\documentclass[onecolumn,aps,pra,preprint,superscriptaddress,longbibliography]{revtex4-1}
\usepackage{amsmath}
\usepackage{graphicx}
\usepackage[lofdepth,lotdepth, caption=false]{subfig}
\usepackage{verbatim}
\usepackage{color}
\usepackage{dcolumn}
\usepackage{bm}
\usepackage{miller}
\usepackage{color}

\usepackage{xr-hyper}
\usepackage{hyperref}

\usepackage[english]{babel}

\begin{document}

\title{Gapless Dirac magnons in CrCl$_{3}$}

\author{John A.~Schneeloch}
\affiliation{Department of Physics, University of Virginia, Charlottesville,
Virginia 22904, USA}

\author{Yu Tao}
\affiliation{Department of Physics, University of Virginia, Charlottesville,
Virginia 22904, USA}

\author{Yongqiang Cheng}
\affiliation{Neutron Scattering Division, Oak Ridge National Laboratory, Oak Ridge, Tennessee 37831, USA}

\author{Luke Daemen}
\affiliation{Neutron Scattering Division, Oak Ridge National Laboratory, Oak Ridge, Tennessee 37831, USA}

\author{Guangyong Xu}
\affiliation{NIST Center for Neutron Research, National Institute of Standards and Technology, Gaithersburg, Maryland 20877, USA}

\author{Qiang Zhang}
\affiliation{Neutron Scattering Division, Oak Ridge National Laboratory, Oak Ridge, Tennessee 37831, USA}

\author{Despina Louca}
\thanks{Corresponding author}
\email{louca@virginia.edu}
\affiliation{Department of Physics, University of Virginia, Charlottesville,
Virginia 22904, USA}

\maketitle

\textbf{Bosonic Dirac materials are testbeds for dissipationless spin-based electronics. In the quasi two-dimensional honeycomb lattice of CrX$_{3}$ (X=Cl, Br, I), Dirac magnons have been predicted at the crossing of acoustical and optical spin waves, analogous to Dirac fermions in graphene. 
Here we show that, distinct from CrBr$_{3}$ and CrI$_{3}$, gapless Dirac magnons are present in bulk CrCl$_{3}$, with inelastic neutron scattering intensity at low temperatures approaching zero at the Dirac $K$ point. Upon warming, magnon-magnon interactions induce strong renormalization and decreased lifetimes, with a $\sim$25\% softening of the upper magnon branch intensity from 5 to 50 K, though magnon features persist well above $T_N$. 
Moreover, an unusual negative thermal expansion (NTE) of the $a$-axis lattice constant and anomalous phonon behavior are observed below 50 K, indicating magnetoelastic and spin-phonon coupling arising from an increase in the in-plane spin correlations that begins tens of Kelvin above $T_N$.}

\section{Introduction}

Two-dimensional (2D) honeycomb lattices often exhibit topological properties, such as the quantum Hall effect, even in the absence of a magnetic field because of quantum confinement \cite{haldane_model_1988}. Graphene is such a system, hosting massless Dirac fermions at the Fermi surface. Bosonic systems can also have topologically nontrivial band structures that give rise to behaviors such as the spin Nernst effect \cite{kim_realization_2016} and a thermal Hall effect \cite{mcclarty_topological_2021}. Included in this list is CrX$_{3}$ (X=Cl,Br,I), consisting of weakly bound van der Waals layers, considered to be the magnetic analogues of graphene. Each layer is a 2D honeycomb lattice of magnetic Cr$^{3+}$ ions with spin $S=3/2$.
The ground state is either ferromagnetic (FM), as in CrBr$_{3}$ and CrI$_{3}$ with an out-of-plane spin orientation, or antiferromagnetic with an in-plane FM alignment that alternates in the perpendicular direction, as in the insulating CrCl$_{3}$.
All of the CrX$_{3}$ compounds exhibit a structural phase transition driven by shifts in the layers from the high-temperature (HT) monoclinic $C2/m$ phase to the low-temperature (LT) ABC rhombohedral $R\bar{3}$ phase (Fig.\ \ref{fig:1}(a)) \cite{mcguire_crystal_2017}, though this transition is often frustrated \cite{mcguire_magnetic_2017}. 
In CrCl$_{3}$ the magnetic easy axis is in-plane (Fig. 2(b)) while in CrBr$_{3}$ and CrI$_{3}$, the easy axis is out-of-plane with stronger interlayer coupling \cite{mcguire_crystal_2017,narath_zero-field_1965}. The interlayer magnetic coupling is stacking-dependent as demonstrated by a tenfold increase of AFM coupling strength reported in a thin $C2/m$-structure-stacked CrCl$_{3}$ \cite{klein_enhancement_2019}. The edge-sharing octahedral coordination of Cr$^{3+}$ with its 3d$^{3}$ electronic configuration provides for superexchange spin-spin interactions. Unlike in the heavier halides, the spin-orbit coupling (SOC) is expected to be much weaker in CrCl$_{3}$.

The magnetic transitions in CrCl$_{3}$ are especially complex. From heat capacity, a single sharp peak corresponding to the FM transitions in CrBr$_{3}$ \cite{yu_large_2019} and CrI$_{3}$ \cite{mcguire_coupling_2015} is evident. For CrCl$_{3}$, however, while a sharp peak is seen at the AFM transition at 14.1 K, an additional broad hump is present at 17.2 K \cite{mcguire_magnetic_2017}. This hump, as well as an inflection point observed in Faraday rotation data at 16.8 K \cite{kuhlow_magnetic_1982}, has been interpreted as signaling the onset of a pseudo-FM phase having in-plane FM order but inter-plane disorder. 
Similar observations have been made for  CrSBr, another layered compound with the same in-plane-FM and inter-plane-AFM order  \cite{lee_magnetic_2021}. AC susceptibility measurements on CrCl$_{3}$ provide additional evidence for two qualitatively different magnetic transitions; the real part of the magnetic susceptibility shows two peaks, located at 14.4 and 16.0 K, but the imaginary part only shows a single peak at 16.0 K \cite{liu_anisotropic_2020}. Finally, we note that one study has reported the possibility of two or three transitions from low-field magnetic susceptibility measurements, as well as a hysteresis in the intensity of a magnetic peak measured by neutron diffraction on warming and cooling \cite{bykovetz_critical_2019}. 

Nuclear magnetic resonance (NMR) \cite{narath_spin-wave_1965,narath_nuclear_1964, davis_spin-wave_1964,gossard_experimental_1961, narath_zero-field_1965} and inelastic tunneling spectroscopy \cite{kim_evolution_2019} have been used to probe the nature of the magnetic interactions in CrX$_{3}$. In CrCl$_{3}$, the nearest-neighbor in-plane exchange constant ($J$) is $J$ = -0.90 meV from NMR in the range 0.4$\leq$T$\leq$8.1 K \cite{narath_spin-wave_1965}, and $J$ = -0.92 meV from inelastic tunneling spectroscopy at 0.3 K \cite{kim_evolution_2019}. 
Interlayer interactions have also been probed by NMR, resulting in an out-of-plane AFM exchange constant of +0.003 meV (for a simplified model) \cite{narath_spin-wave_1965}, and an AFM resonance \cite{macneill_gigahertz_2019, kapoor_observation_2021}. Complementary to these techniques is inelastic neutron scattering (INS), as it can probe magnetic excitations across a wide range of wavevectors $\mathbf{Q}$. CrBr$_{3}$ \cite{samuelsen_spin_1971,yelon_renormalization_1971,cai_topological_2021} and CrI$_{3}$ \cite{chen_topological_2018} have been probed by INS and both have been proposed to host topological magnons \cite{chen_topological_2018, cai_topological_2021}. Very recently, CrCl$_{3}$ has been measured by INS \cite{chen_massless_2021}, but data from only a single temperature (4 K) has been reported. Studying a wide range of temperatures would yield insights on the nature of the magnetism in CrCl$_{3}$.

We report elastic and inelastic neutron scattering measurements of the temperature dependence of the magnon and phonon excitations in CrCl$_{3}$. Gapless Dirac magnons are observed at the lowest temperature where the dispersion intensity drops to zero at 4.5 meV, at the intersection of the optical and acoustic branches. On warming, the lower magnon branch shows a weak temperature dependence while the upper magnon branch shows a strong energy renormalization and lifetime broadening that is indicative of strong magnon-magnon interactions. Furthermore, the negative thermal expansion (NTE) of the honeycomb plane that sets in below $\sim$50 K along with signs of spin-phonon coupling are features of magnetoelastic effects in CrCl$_{3}$ that suggest the onset of in-plane FM spin correlations well above the Néel temperature.

\section{Results and Discussion}

\begin{figure}[ht]
\begin{center}
\includegraphics[width=16cm]
{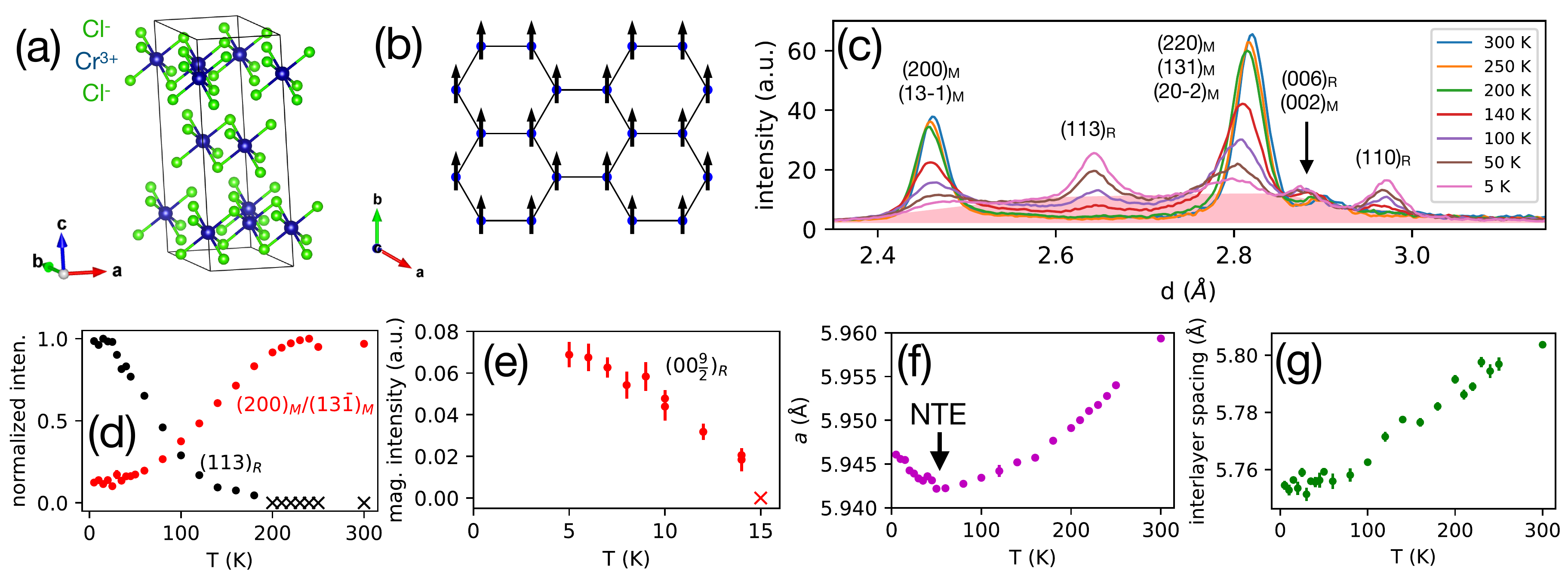}
\end{center}
\caption{(a) The crystal structure diagram of the $R\bar{3}$-phase in CrCl$_{3}$. (b) A honeycomb layer in CrCl$_{3}$. (c) Powder diffraction data (intensity vs.\ lattice spacing $d$) collected on cooling from 300 to 5 K with Bragg peaks belonging to the two phases, labeled M-monoclinic and R-rhombohedral. The shaded region shows the diffuse scattering part of the intensity at 5 K. (d) Integrated intensity of the $(113)_R$ and $(200)_M$/$(13\bar{1})_M$ peaks is plotted as a function of temperature, showing the progression of the transition on cooling from 300 K. Error bars smaller than symbols. (e) Integrated intensity of the magnetic $(00\frac{9}{2})_R$ peak, present below 15 K. Data collected on both cooling and warming. (f) Temperature-dependence of the $a$-axis lattice constant. Negative thermal expansion (NTE) is observed below 50 K. (g) Interlayer spacing obtained from the position of the $(002)_M$/$(006)_R$ Bragg peak. Data in (f) and (g) collected on cooling.}
\label{fig:1}
\end{figure}

\begin{figure}[ht]
\begin{center}
\includegraphics[width=16cm]
{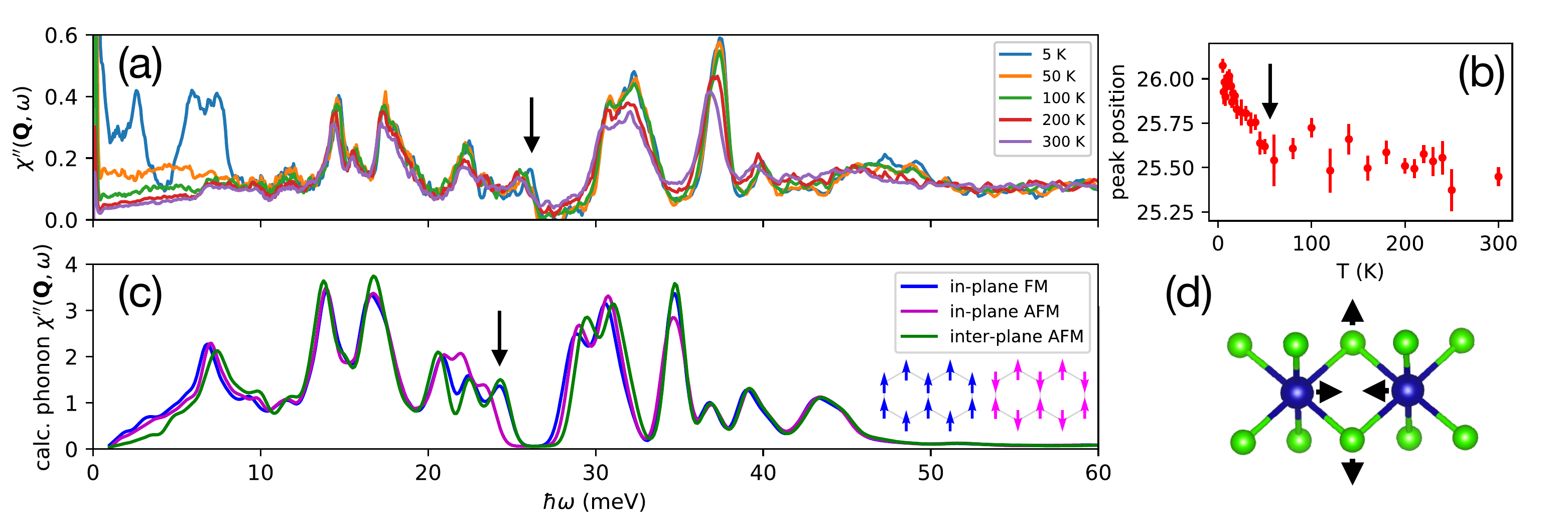}
\end{center}
\caption{(a) A plot of the dynamic susceptibility, $\chi^{\prime \prime} (\mathbf{Q}, \omega)$ as a function of temperature along the high-$Q$ trajectory. Intensities have been smoothed; for clarity, errorbars not shown. (b) Position of the 26 meV phonon feature from fitting. (c) DFT-calculated phonon intensity along the high-$Q$ trajectory for the $R\bar{3}$ structure with spins fixed to a certain ordering, including FM, in-plane AFM, or inter-plane AFM (i.e., that observed in CrCl$_{3}$). Inset depicts in-plane FM or AFM alignment. The features in the DFT intensity are about 7\% lower in $\hbar \omega$ than the corresponding features in (a). (d) Atomic movement of the 26 meV phonon mode.}
\label{fig:2}
\end{figure}

Neutron scattering measurements were performed on a CrCl$_{3}$ powder sample at the VISION instrument at Oak Ridge National Laboratory, a time-of-flight spectrometer that simultaneously measures elastic and inelastic scattering data. The elastic intensity is shown in Fig.\ \ref{fig:1}(c) as a function of temperature. Marked on the plot are Bragg peaks corresponding to the monoclinic $C2/m$ and rhombohedral $R\bar{3}$ phases. It can be seen that the $C2/m$$\rightarrow$$R\bar{3}$ transition is frustrated. Starting from 300 K, the structure is essentially single phase $C2/m$, and all Bragg peaks can be indexed as such. Upon cooling, a gradual transition to the rhombohedral structure is observed, accompanied by diffuse scattering that most likely arises from a disordered sequence of HT- and LT-type layer stacking. 
Shown in Fig.\ \ref{fig:1}(d) is the temperature dependence of the integrated intensity of the $(113)_R$ peak and the overlapping $(200)_M$/$(13\bar{1})_M$ peaks. Coexistence of both phases is observed over a wide temperature range in the powder sample starting at $\sim$200 K and ending at $\sim$25 K, that is distinctly different from the single crystal results reported earlier \cite{mcguire_magnetic_2017}. 
The transition proceeds gradually below 200 K until it stops around 25 K. Below $\sim$15 K, the $(0,0,9/2)_R$ AFM Bragg peak is observed at $d$-spacing $\sim$3.84 \AA, consistent with the anticipated onset of AFM order at the Néel temperature (Fig.\ \ref{fig:1}(e)). Thus, the magnetic and structural transitions are not coupled. 

An unusual behavior, NTE, is observed in the $a$-axis lattice constant on cooling (Fig.\ \ref{fig:1}(f)), which appears not to be associated with stacking changes. Since in-plane layer translations accompanying the LT- or HT-type stacking are nearly commensurate, Bragg peaks with $H_R$ and $K_R$ Miller indices divisible by 3 will be largely unaffected by variations in LT-type or HT-type stacking. (See Supplement for mathematical details.) For illustration, Fig.\ S1 in the Supplement shows the $(H0L)$ plane of single-crystal X-ray scattering measurements; although the structural transition is frustrated, the peaks along $(\bar{6}0L)$ are undisturbed. Further cooling locks in the rhombohedral structure but with substantial diffuse scattering, evident underneath the Bragg peaks. Similar diffuse scattering has been noted for $\alpha$-RuCl$_{3}$ \cite{johnson_monoclinic_2015}. 

The NTE behavior is observed below 50 K. The temperature dependence of the $a$-axis lattice constant obtained from the position of the $(300)_R$ Bragg peak is shown in Fig.\ \ref{fig:1}(f). The intensity near the $(300)_R$ peak is shown in Fig.\ S2(a) in the Supplement. Data collected on powder and single-crystal CrCl$_{3}$ on POWGEN (ORNL) and SPINS (NCNR), respectively, show the same temperature dependence in $a$; see Fig.\ S3 in the Supplement. A similar lattice constant anomaly has been observed in other quasi-2D honeycomb magnets, namely CrBr$_{3}$ \cite{kozlenko_spin-induced_2021} and Cr$_{2}$Ge$_{2}$Te$_{6}$  \cite{carteaux_crystallographic_1995}, though in those materials the anomaly occurs near the magnetic ordering temperatures. The ilmenite compound NiTiO$_{3}$, which shares the quasi-2D magnetic structure of CrCl$_{3}$ but with 3D structural bonding, also exhibits an anomaly in its lattice constants \cite{sauerland_magnetostructural_2021}. In the out-of-plane direction, the interlayer spacing obtained from the $(002)_M$/$(006)_R$ peak position plotted in Fig.\ \ref{fig:1}(g) shows no clear anomaly which suggests that the NTE behavior is confined in the honeycomb plane.

The NTE behavior coincides with changes in the phonon inelastic spectrum below $\sim$50 K, most prominently near 26 meV (as first seen via Raman spectroscopy \cite{glamazda_relation_2017}.) In our experiment, the INS intensity was measured along two narrow paths in ($Q$, $\omega$) space, which are labeled the low-$Q$ and high-$Q$ trajectories in Fig.\ 3(a). Intensity maps of $\chi^{\prime \prime} (\mathbf{Q},\omega)$ at temperatures from 5 to 300 K can be found in the Supplemental Materials. 
In Fig.\ \ref{fig:2}(a), the dynamic susceptibility $\chi^{\prime \prime} (\mathbf{Q}, \omega)$ along the high-$Q$ trajectory is plotted at select temperatures. Several changes across the phonon spectra are observed with temperature, including the one at 26 meV indicated with an arrow. Shown in Fig.\ 2(b) is the fitted position of the 26 meV phonon peak as it rapidly shifts to higher energies upon cooling, starting around $\sim$50 K and coinciding with the NTE. The 26 meV mode is a Cr-Cr stretching mode in the honeycomb plane (Fig.\ \ref{fig:2}(d).) We performed density functional theory (DFT) calculations using three different magnetic structures: in-plane FM, in-plane AFM, or the observed inter-plane AFM, all of which have spins oriented in-plane. Slightly below 26 meV in the calculated spectra in Fig.\ 2(c), an anomalous decrease in the phonon energies is seen for in-plane AFM alignment relative to either in-plane FM or inter-plane AFM, showing that inter-plane order has little effect on the 26 meV phonon anomaly compared to the in-plane order. 
Since motion from the 26 meV mode involves Cr ion oscillations, this mode may induce frustration to the spin correlations.
The differences in the relaxed lattice constants in the DFT calculations match the magnetoelastic coupling trends noted earlier, with $a$ larger and $c$ smaller in the FM alignment as compared to the in-plane AFM alignment, and little difference seen between FM and inter-plane AFM (see Table S1 in the Supplement.) These calculations confirm that an increase in the in-plane spin-spin correlation on cooling, combined with the presence of spin-phonon coupling, is the cause of the phonon anomaly. It is remarkable that this substantial increase in spin correlations occurs well above the ordering temperature, on the order of $50$ K $\sim 3 T_N$.


\begin{figure}[ht]
\begin{center}
\includegraphics[width=16cm]
{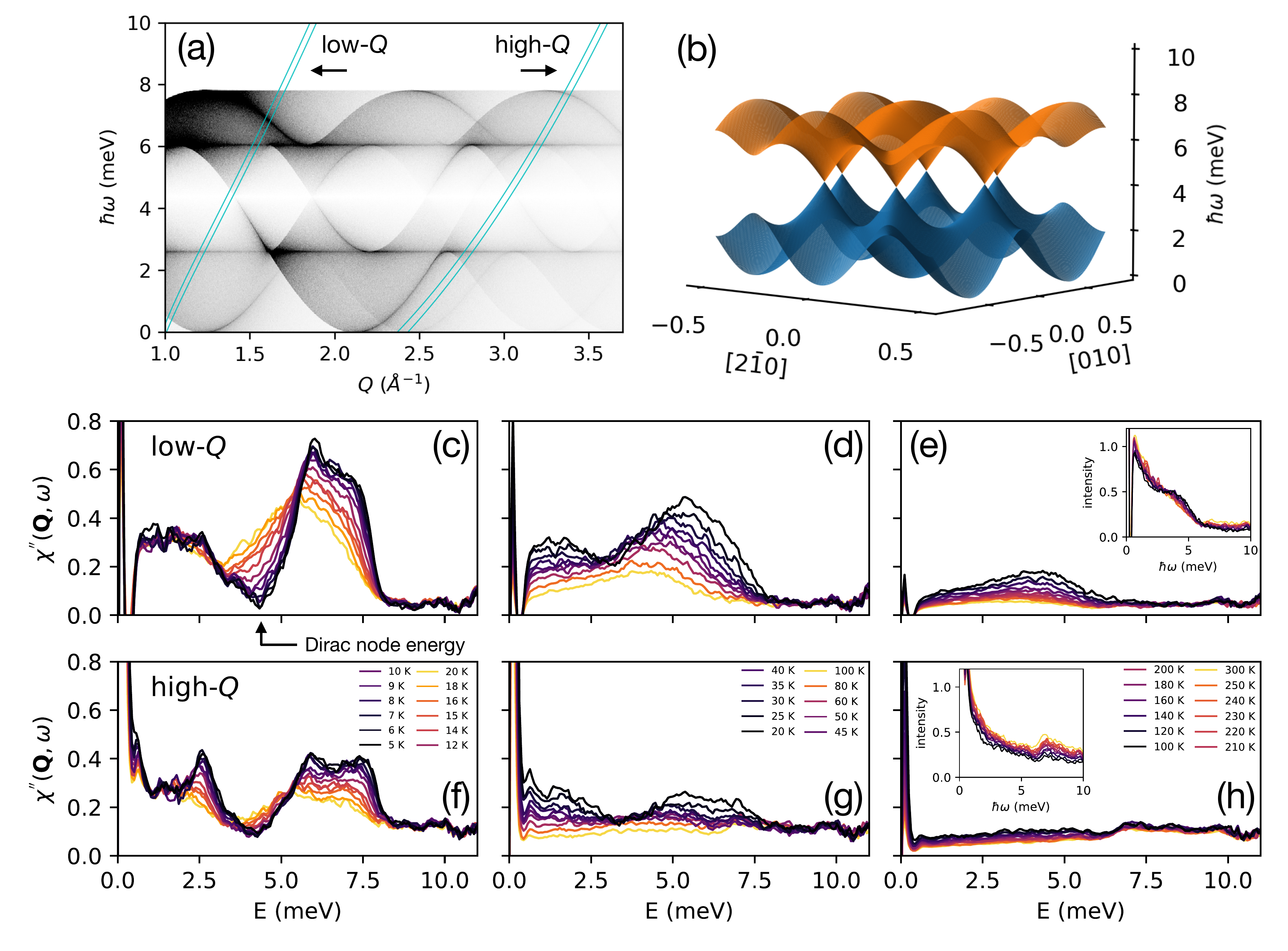}
\end{center}
\caption{(a) Powder-averaged $S_{\perp}(\mathbf{Q},\omega)$ simulated via SpinW, with low- and high-$Q$ trajectories depicted in light blue. 
(b) Spin-wave dispersion in the $(HK0)_R$ plane from fitted magnetic exchange coefficients. 
(c-h) Bose-factor-corrected inelastic neutron scattering intensity in the temperature ranges of (a,d) 5-20 K, (b,e) 20-100 K, and (c,f) 100-300 K, along the (a-c) low-$Q$ and (d-f) high-$Q$ trajectories. Insets of (c) and (f) show the uncorrected intensities of the data in those panels. All intensities smoothed, and errorbars not included for clarity.}
\label{fig:3}
\end{figure}

\begin{figure}[ht]
\begin{center}
\includegraphics[width=16cm]
{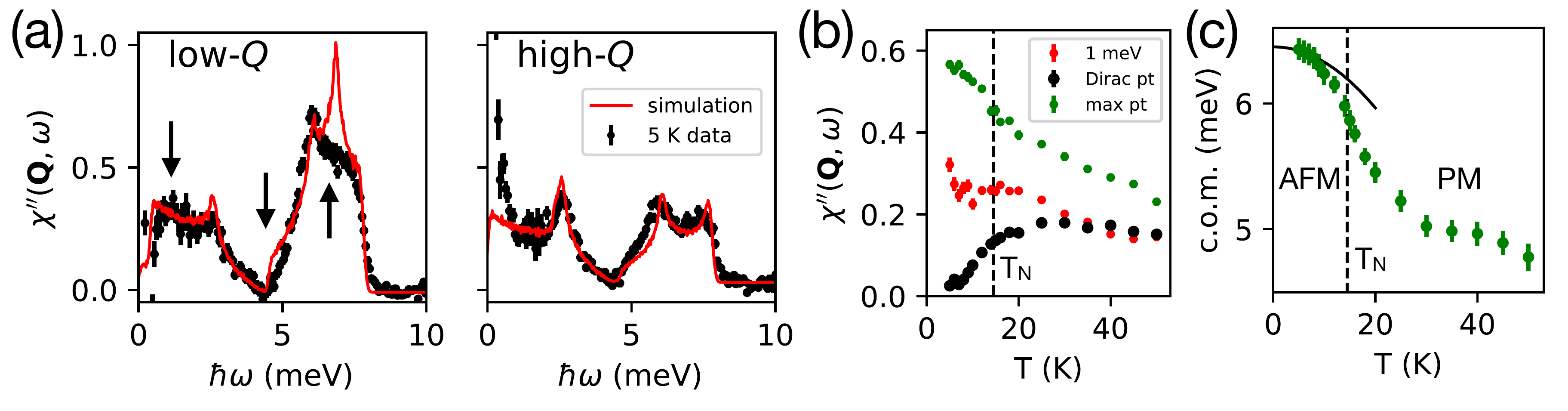}
\end{center}
\caption{(a) Comparison of 5 K inelastic data with calculated $\chi^{\prime \prime} (\mathbf{Q}, \omega)$ along low- and high-$Q$ trajectories. The 5 K data has the 300 K data subtracted to eliminate the contribution from the 7-10 meV phonon feature, which can be seen in, e.g., the high-$Q$ data in Fig.\ \ref{fig:3}(f). Data are unsmoothed. (b) Intensities near 1 meV, the Dirac node energy, and the middle of the upper branch (``max pt'') for the low-$Q$ trajectory. The Dirac node energies and the position and energy of the midpoint of the upper branch hump were determined by fits to a function of connected line segments. The intensities near the Dirac node and 1 meV were taken as an average within $\pm0.1$ meV of these locations. (c) The center-of-mass (c.o.m.) position of the upper branch intensity plotted as a function of temperature. The black curve shows the expected first-order renormalization $\hbar \omega = \hbar \omega_0 (1 - \alpha_1 T^2)$ of the honeycomb ferromagnet with nearest-neighbor interactions, where $\alpha_1 = \pi / (24 \sqrt{3} J^2 S^3)$ \cite{pershoguba_dirac_2018}.
}
\label{fig:4}
\end{figure}


The characteristic magnon spectrum of CrCl$_{3}$ manifests as a drastic change in the low-energy $\chi^{\prime \prime} (\mathbf{Q}, \omega)$ data between 5 and 50 K, as seen in Fig.\ 2(a). 
The temperature changes of the magnon intensity are shown in detail in Figures \ref{fig:3}(c-h). A very strong temperature dependence is observed along both the low- and high-$Q$ trajectories. To explain the observed patterns, the spin waves were modeled using a Heisenberg Hamiltonian for a 2-dimensional honeycomb lattice with FM nearest-neighbor (n.n.) interactions, which, analogous to graphene, features Dirac nodes at the $K$ and $K^{\prime}$ points of the Brillouin zone \cite{pershoguba_dirac_2018}, though we added coupling for 2nd- and 3rd-nearest in-plane neighbors as well. 
The dispersion is shown in Fig.\ \ref{fig:3}(a). Due to the very weak interlayer magnetic coupling in CrCl$_{3}$ \cite{narath_spin-wave_1965}, we presume the dispersion is flat along the $L$-direction. 
(In the Supplement (Fig.\ S4), we compare simulated magnon intensity for an interlayer coupling (per nearest interlayer bond) of +0.04 meV for the $R\bar{3}$ and $C2/m$ structures to zero interlayer coupling, and find little difference. The comparison suggests that a relatively large $J_L$ value of 0.04 meV is not enough to qualitatively change the intensity, and that the influence of stacking disorder on the magnon intensity is negligible.) 
The calculated powder-averaged spin correlation function $S_{\perp}(Q,W)$ is shown in Fig.\ \ref{fig:3}(a) \cite{toth_linear_2015}. The paths of the low- and high-$Q$ trajectories are plotted as light blue curves, with the separation corresponding to the full-width-at-half-maximum of the energy resolution. 
The integrated intensity along dispersion's maximum, saddle points, and Dirac nodes leave sharply-defined features in the intensity along the low- and high-$Q$ trajectories which have $Q$-independent energies (due to the flat dispersion along $L$). 
From the energies of the Dirac node, saddle points, and dispersion maximum, we extracted magnetic coupling constants of $J_1 = -0.940(5)$, $J_2 = -0.035(2)$, and $J_3=+0.070(4)$ meV. 
The uncertainties (from least-squares fitting) are likely underestimates, given the imperfect fit of the model to the data (see below); nevertheless, the obtained $J_1$ values roughly agree with estimates from NMR \cite{narath_spin-wave_1965} and inelastic tunneling spectroscopy \cite{kim_evolution_2019}.

We compare the simulated intensity with the 5 K inelastic data in Fig.\ \ref{fig:4}(a). The Dirac node at 4.5 meV, nestled between the upper and lower magnon dispersion branches where the intensity goes to zero, is clearly seen in the data. Note that the Dirac node is gapless (a finding also reported in a recent INS study \cite{chen_massless_2021}), suggesting the absence of Kitaev or Dzyaloshinskii-Moriya interactions. The dispersion saddle points result in peaks near 2.5 and 6 meV. The band maximum manifests as a steep dropoff in intensity near 8 meV. Although the data agree reasonably well with the $J_1$-$J_2$-$J_3$ Heisenberg model, there are discrepancies, most notably a model peak near 7 meV at low $Q$ which is not seen in our data. This peak corresponds to the crossing of the low-$Q$ trajectory with a boundary (see Fig.\ \ref{fig:3}(a)) corresponding to where new magnon branches become accessible as $Q$ changes at constant $\hbar \omega$. We speculate that, even at 5 K, magnon-magnon interactions might cause disorder in the spin orientations that suppress this peak. More studies are needed to explore this issue further.

On warming, the dispersion features soften and broaden (as seen in Fig.\ \ref{fig:3}(c-h)), but many features remain well-defined above $T_N$. The ``corner'' near 6 meV at low-$Q$ remains visible at 20 K, and a ``hump'' of upper branch intensity is still present at 60 K. The valley near 4.5 meV shifts to lower frequency due to magnon energy renormalization, indicating that spectral weight above and below the Dirac node fills in the surrounding energy range. The upper edge near 8 meV, already broader than resolution at 5 K (Fig.\ \ref{fig:4}(a)), becomes markedly less steep on warming. Notably, the peak near 2.5 meV, especially large at high-$Q$, vanishes around $T_N$ after shrinking steadily with temperature. 

To characterize the temperature changes in the magnon intensity, the integrated intensity at three points of the 5 K data in Fig.\ \ref{fig:4}(a) (marked with arrows) are plotted as a function of temperature in Fig.\ \ref{fig:4}(b). 
These points are 1 meV, the Dirac node, and ``max pt'' (the midpoint between the two corners of the upper branch intensity, which are $\sim$6 and 7.5 meV in Fig.\ \ref{fig:4}(a)), with the latter two points defined via a fit to a function of connected line segments (see Supplement). The data were integrated within $\pm$0.1 meV of these points. Starting with the Dirac node, the intensity is $\sim$zero at 4.5 meV, but gradually increases because of intensity spilling over from the upper branch due to the magnon energy renormalization. In contrast, the intensities at the other two points, representative of the lower and upper magnon branches, decrease with temperature, though, below $T_N$, the lower branch changes more gradually than the upper branch. In Fig.\ \ref{fig:4}(c), we show the center-of-mass (c.o.m.) position of the upper branch intensity. It turns out that the higher energy branch softens more rapidly than can be accounted for by the first-order magnon-magnon renormalization for a honeycomb ferromagnet \cite{pershoguba_dirac_2018}, shown as the black curve. Thus, the magnons in CrCl$_{3}$ exhibit significant broadening in energy, and a shift to lower energies, as shown by the $\sim$2 meV ($\sim$25 \%) decrease in the c.o.m.\ position from 5 to 50 K.
As noted, even at 5 K we observe a broadening that is greater than resolution, in contrast to a recent study (on an instrument with lesser resolution) which was not able to observe such broadening at 4 K \cite{chen_massless_2021}. Our data at a multitude of additional temperatures further emphasize the substantial broadening of the magnon features on warming.

At high temperature, we expect the spin waves to become overdamped, leading to quasi-elastic scattering with a Lorentzian lineshape, and no change with temperature except if the magnon decay rate or $\mathbf{Q}$-coefficients change \cite{zaliznyak_magnetic_2004}. In the insets of Figures \ref{fig:3}(e,h), we see that the intensity (without Bose factor correction) is roughly constant from 100 to 300 K, especially for the low-$Q$ trajectory where phonons have less contribution. This temperature-independent intensity is consistent with expectations, provided that the magnon decay rate is temperature-independent. In the Supplement (Fig.\ S7), we show that the overall intensity ($\chi^{\prime \prime} (\mathbf{Q}, \omega)$ integrated from 1 to 10 meV) decreases more rapidly at high-$Q$ than low-$Q$, especially at low temperature. 


The onset of spin waves occurs on the same temperature scale as the in-plane FM spin correlations, as indicated via the anomalies in the $a$-axis lattice constant and the phonon feature; these behaviors all occur gradually on cooling below $\sim$50 K, 10s of K higher than $T_N$. In contrast, CrBr$_{3}$ exhibits abrupt changes at the Curie temperature in the spin-phonon coupling and the lattice constants \cite{kozlenko_spin-induced_2021}, and these differences may be due to differing magnetic anisotropy. Anisotropy can play an important role in the ordering behavior of 2D systems; the 2D isotropic Heisenberg model has no long-range order above $T>0$ for short-range interactions \cite{mermin_absence_1966}, in contrast to the transition present in the 2D Ising model \cite{onsager_crystal_1944}. The different spin orientations of the chromium trihalides hints at a different magnetic anisotropy for CrCl$_{3}$ than for CrBr$_{3}$ and CrI$_{3}$. Though estimates from inelastic tunneling spectroscopy suggest that the anisotropy is on the order of a few percent or less \cite{kim_evolution_2019}, even a slight asymmetry may influence magnetic behavior, such as the presence of a Kosterlitz-Thouless transition in K$_{2}$CuF$_{4}$ \cite{hirakawa_neutron_1982}.
Studying the magnetic excitations of CrCl$_{3}$ in detail may elucidate the role of anisotropy in the magnetism of quasi-2D materials. 

A diverse array of materials have been investigated for their topological properties; CrBr$_{3}$ \cite{cai_topological_2021} and CrI$_{3}$ \cite{chen_topological_2018} have both been reported to host topological magnons, due to the Dzyaloshinskii-Moriya interaction opening energy gaps at the Dirac points. In contrast, our CrCl$_{3}$ data show a sharp cusp at the Dirac point in powder-averaged INS intensity, indicating no gap within the precision of our experiment. CrCl$_{3}$ thus provides a foil to the other chromium trihalides, and may facilitate understanding of spin dynamics across the CrX$_{3}$ system. Meanwhile, the remarkable temperature dependence of CrCl$_{3}$, in which the magnon intensity has only subtle changes across $T_N$ and continues to have identifiable upper branch intensity tens of Kelvin above $T_N$, suggests that the temperature dependence of CrBr$_{3}$ and CrI$_{3}$ should be investigated in more detail, especially above T$_C$. For CrI$_{3}$, on approaching $T_C$, the magnetic anisotropy gap appears to vanish while the acoustic magnon stiffness remains substantial \cite{chen_topological_2018}, but data well above $T_C$ has not been reported and would be informative. Further study should clarify the extent to which the behavior of CrCl$_{3}$ is unusual. 

In conclusion, in our inelastic neutron scattering measurements we observed changes in the magnon and phonon intensities. At 5 K, the magnon intensity shows a sharp cusp indicating a dispersion with gapless Dirac magnons. On warming, a substantial energy renormalization is seen that occurs gradually on warming, with little discontinuity at $T_N$ and upper branch intensity still visible tens of Kelvin above $T_N$. Below $\sim$50 K, a negative thermal expansion and an increase in the energy of a phonon feature near 26 meV can be seen. These anomalies are explained via DFT calculations as resulting from spin-phonon and magnetoelastic coupling combined with the increase in in-plane spin correlations on cooling. Altogether, our results suggest a persistence of in-plane spin correlations to the order of $\sim$3 $T_N$, as seen from the behavior of the $a$-axis lattice constant, phonons, and magnons.

\nocite{morosin_xray_1964,johnson_monoclinic_2015,kozlenko_spin-induced_2021,narath_spin-wave_1965,klein_enhancement_2019,seeger_resolution_2009}

\section*{Methods}
Most data were collected on VISION \cite{seeger_resolution_2009}, an indirect-geometry time-of-flight neutron spectrometer at the Spallation Neutron Source (SNS) at Oak Ridge National Laboratory (ORNL). The final neutron energy was fixed at 3.5 meV. Inelastic data were taken on two detector banks at scattering angles of 45 and 135$^{\circ}$, while six elastic scattering detector banks were located at 90$^{\circ}$. The CrCl$_{3}$ powder was poured into a vanadium can. Background from an empty-can scan was subtracted from all data. Measurements were also taken at the time-of-flight instrument POWGEN \cite{huq_powgen_2019} (at the SNS at ORNL) and on the triple-axis spectrometer SPINS (at the NIST Center for Neutron Research.)

The powder samples measured in VISION and POWGEN consisted of as-purchased CrCl$_{3}$ powder from Alfa Aesar (99.9\% purity, metals basis), consisting of mm-sized flakes. The single crystal measured on SPINS was grown via the vapor transport growth procedure detailed in Ref.\ \cite{mcguire_magnetic_2017}. 

Spin-polarized Density Functional Theory (DFT) calculations of CrCl$_{3}$ (in its $C2/m$ and $R\bar{3}$ phases and with FM in-plane AFM, and inter-plane AFM configurations) were performed using the Vienna Ab initio Simulation Package (VASP) \cite{kresse_efficient_1996}. 
The calculation used the Projector Augmented Wave (PAW) method \cite{blochl_projector_1994,kresse_ultrasoft_1999} to describe the effects of core electrons, and the Perdew-Burke-Ernzerhof (PBE) \cite{perdew_generalized_1996} implementation of the Generalized Gradient Approximation (GGA) for the exchange-correlation functional. 
The energy cutoff was 600 eV for the plane-wave basis of the valence electrons. The lattice parameters and atomic coordinates from Ref.\ \cite{morosin_xray_1964} were used as the initial structure, and they were then fully relaxed to minimize the potential energy. The electronic structure was calculated on a $\Gamma$-centered mesh (9×5×9 for $C2/m$ and 9×9×3 for $R\bar{3}$.) 
The total energy tolerance for electronic energy minimization was 10$^{-8}$ eV, and for structure optimization it was 10$^{-7}$ eV. The maximum interatomic force after relaxation was below 0.001 eV/\AA. The optB86b-vdW functional \cite{klimes_chemical_2010} for dispersion corrections was applied, and a Hubbard U term of 3.7 eV \cite{wang_oxidation_2006} was applied to account for the localized 3d orbitals of Cr. 
A supercell ($2 \times 1 \times 2$  for $C2/m$ and $2 \times 2 \times 1$ for $R\bar{3}$) was created for phonon calculations. The interatomic force constants were calculated by Density Functional Perturbation Theory (DFPT), and the vibrational eigenfrequencies and modes were then calculated using phonopy \cite{togo_first_2015}. The OCLIMAX software \cite{cheng_simulation_2019} was used to convert the DFT-calculated phonon results to the simulated INS spectra.

A standard deviation of statistical uncertainty is denoted with parentheses in the last digit(s) for numbers, and with error bars for plots.

\section*{Acknowledgements}

This work has been supported by the Department of Energy, Grant number
DE-FG02-01ER45927. A portion of this research used resources at the High Flux Isotope Reactor and the Spallation Neutron Source, which are DOE Office of Science User Facilities operated by Oak Ridge National Laboratory. 
Computing resources for DFT simulations were made available through the VirtuES and the ICE-MAN projects, funded by Laboratory Directed Research and Development program and Compute and Data Environment for Science (CADES) at ORNL.
We acknowledge the support of the National Institute of Standards and Technology, US Department of Commerce, in providing neutron research facilities used in this work. 
Certain commercial materials are identified in this paper to foster understanding. Such identification does not imply recommendation or endorsement by the National Institute of Standards and Technology, nor does it imply that the materials or equipment identified are necessarily the best available for the purpose.

\section*{Author Contributions}
J.~S.\ and Y.~T.\ took part in the neutron scattering experiments and analyzed the data. Y.~C., L.~D., G.~X., and Q.~Z.\ helped us carry out our experiments as instrument scientists at the neutron scattering instruments. D.~L.\ conceived of and supervised the project. J.~S.\ and D.~L.\ wrote the manuscript.


%

\end{document}